# Mechanistic Insight into BEOL Thermal Transport via Optical Metrology and Multiphysics Simulation

Yang Shen[1], Shangzhi Song[1], Tao Chen[1], Kexin Zhang[1], Yu Chen[2], Lu Zhao[2], Puqing Jiang[1,*]

[1]*School of Energy and Power Engineering, Huazhong University of Science and Technology, Wuhan, Hubei 430074, PR China*
[2]*Huawei Technologies Co. Ltd., Beijing 100085, PR China*

**Abstract:** As integrated circuits continue to scale down and adopt three-dimensional (3D) stacking, thermal management in the back-end-of-line (BEOL) has emerged as a critical design constraint. In this study, we present a combined experimental and simulation framework to quantitatively characterize and mechanistically understand thermal transport in BEOL multilayers. Using the Square-Pulsed Source (SPS) method, a time-resolved optical metrology technique, we measure cross-plane thermal resistance and areal heat capacity in semiconductor chips at nanometer resolution. Two fabricated chip samples, polished to the M4 and M6 interconnection layers, are analyzed to extract thermal properties of distinct multilayer stacks. Results show that thermal resistance follows a series model, while areal heat capacity scales linearly with metal content. To uncover the underlying physical mechanisms, we perform finite element simulations using COMSOL Multiphysics, examining the influence of via connectivity and dielectric thermal conductivity on effective cross-plane heat transport. The simulations reveal that dielectric materials, due to their large volume fraction, are the primary limiting factor in BEOL thermal conduction, while the via structure plays a secondary but significant role. This combined experimental–simulation approach provides mechanistic insight into heat transport in advanced IC architectures and offers practical guidance for optimizing thermal pathways in future high-performance 3D-stacked devices.

---

[*] Corresponding author
E-mail addresses: jpq2021@hust.edu.cn (P. Jiang)

# 1. Introduction

The rapid advancement of integrated circuits (ICs), driven by Moore's Law and the shift toward three-dimensional (3D) chip stacking, has dramatically increased power density and functional integration [1]. While 3D architectures improve performance and compactness, they also introduce critical challenges in thermal management, particularly along the vertical heat conduction paths [2, 3]. One of the most significant bottlenecks arises from the back-end-of-line (BEOL) interconnect stack, which can exceed 10 μm in thickness and features intricate multilayer geometries composed of metal interconnects and low-thermal-conductivity dielectrics.

The out-of-plane thermal conductivity of BEOL stacks is severely limited by the low-k dielectric materials, which dominate the volume and interrupt efficient heat flow. This makes accurate characterization and modeling of BEOL thermal properties essential for thermal-aware design, simulation, and reliability evaluation in 3D ICs.

Prior efforts to address this challenge include both experimental and modeling approaches. Finite element modeling (FEM) has been widely used to analyze thermal profiles and assess the impact of design parameters such as via density, dielectric choice, and interconnect layout [4-9]. Studies by Fujitsu Limited [5, 6], IMEC [7], and IBM [10] used analytical and steady-state methods to probe BEOL thermal resistance, providing valuable insights into thermal gradients and the effects of material and layout configurations. However, these methods often require custom-fabricated dummy structures and embedded heaters or sensors, limiting their spatial resolution and scalability.

Optical thermoreflectance techniques, such as time-domain thermoreflectance (TDTR) and frequency-domain thermoreflectance (FDTR), offer high spatial and temporal resolution and have advanced our ability to probe nanoscale thermal properties [11, 12]. However, these methods face challenges such as limited measurement range, complex system requirements, and high operational costs. For instance, TDTR's modulation frequency range is limited to 0.1-10 MHz, making it difficult to measure in-plane thermal conductivities below $6\,\text{W/(m·K)}$ [13]. Additionally, TDTR often requires prior knowledge of the sample's specific heat capacity to determine thermal conductivity. FDTR, on the other hand, necessitates meticulous phase correction to ensure data reliability [14, 15].

To overcome these challenges, we present a combined experimental–simulation framework for the mechanistic and quantitative investigation of BEOL thermal transport. At the core of this approach is the Square-Pulsed Source (SPS) technique [16, 17], a time-resolved optical metrology method that enables high-resolution extraction of cross-plane thermal resistance and areal heat capacity without requiring phase data. This study applies SPS measurements to two real semiconductor samples, exposed at different interconnect levels (M4 and M6), to characterize the thermal behavior of individual BEOL segments.

To interpret the measurements and explore structure-property relationships, we complement the experimental results with finite element simulations using COMSOL Multiphysics. These simulations examine how dielectric thermal conductivity and via connectivity govern effective vertical heat conduction in BEOL stacks. Together, the experimental and simulation insights provide a deeper understanding of thermal bottlenecks in advanced chip architectures and inform strategies for thermal pathway optimization in 3D-stacked ICs.

## 2. Sample Structures and Experimental Methodology Overview

### 2.1. Sample Structures

The schematic of the cross-sectional structure of the measured samples is shown in Figure 1(a). Two identical chip samples were polished from the top, with one polished until the M4 layer in the BEOL structure was exposed, and the other polished until the M6 layer was exposed. An aluminum (Al) film was then deposited on their surfaces to serve as a transducer layer for thermoreflectance measurements. These two samples are referred to as the #M4 and #M6 samples, respectively.

The structure of the #M4 sample, as depicted in the thermal model, consists of the following layers from bottom to top: a silicon substrate, a 200 nm thick doped interface layer, a 400 nm thick M1-M4 interconnection layer, and a 100 nm thick Al transducer layer. The #M6 sample structure is similar, with the only difference being that the interconnection layer extends from M1 to M6, with a total thickness of 700 nm. Note that these thicknesses are approximate, and the actual values—considered confidential—were used for data processing in this study.

As labeled in Figure 1(a), M1, M2, M3, M4, M5, and M6 represent the metal interconnection layers with high metal content, while V1, V2, V3, V4, and V5 represent the via layers, which have a lower metal content. Despite their lower metal content, the via layers connect the upper and lower metal interconnection layers, facilitating vertical heat transfer. Therefore, the arrangement of the vias significantly influences the local thermal properties of the nanofilms. By measuring the thermal properties of the BEOL films at various locations, this study systematically investigates the effects of metal content and via positioning on the thermal properties, providing crucial support for the optimized design of chip thermal management.

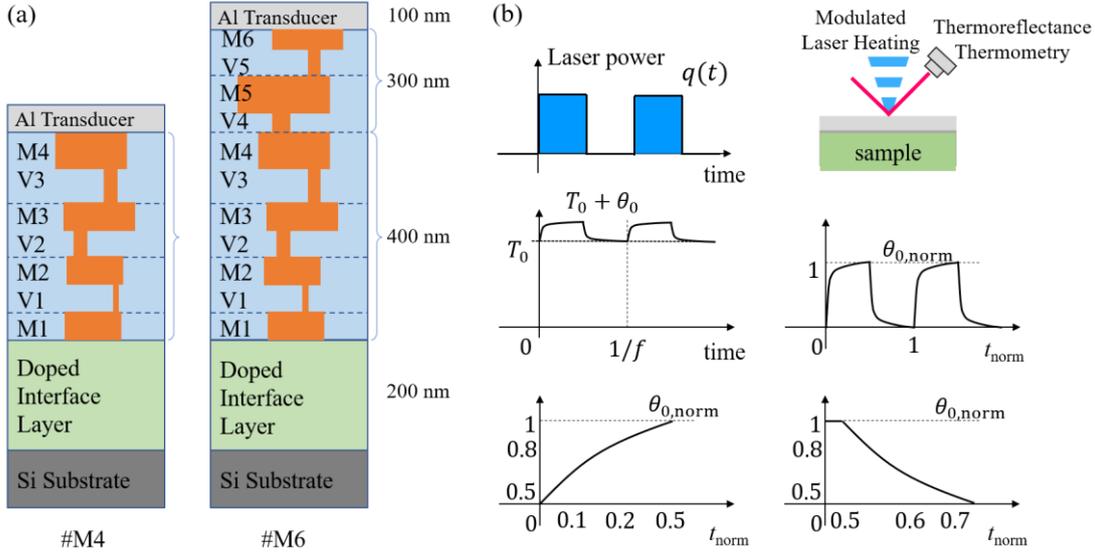

**Fig. 1** (a) BEOL sample structure information. (b) Principle of the SPS method for thermal characterization.

*2.2. Basics of the SPS Method*

The Square-Pulsed Source (SPS) method is a time-resolved, pump–probe optical technique designed for precise thermal characterization of multilayer thin films. As illustrated in Fig. 1(b), the method uses a 50% duty cycle square-wave laser modulation to periodically heat the sample (pump), while a continuous-wave laser measures the resulting surface temperature variations via thermoreflectance (probe). A waveform analyzer captures the full time-domain thermal response over each heating cycle. To extract thermal properties, the measured signals are normalized in both amplitude and time and fitted to thermal model predictions.

The SPS method is particularly well-suited for multilayer structures. Key parameters that influence the thermal response signal include: in-plane thermal conductivity ($k_r$), cross-plane thermal conductivity ($k_z$), volumetric heat capacity ($C$), and thickness ($h$) of each layer, as well as the interfacial thermal conductance ($G$) and laser spot size ($r_0$). These parameters are not independent. Prior work by Chen and Jiang [18, 19] identified several governing combinations that dictate the measured thermal response, such as $\frac{k_r}{Cr_0^2}$, $\frac{\sqrt{k_z C}}{h_m C_m}$, and $\frac{G}{h_m C_m}$, where $h_m C_m$ represents the areal heat capacity of the metal transducer layer. These factors collectively describe how heat diffuses through layered structures and interfaces.

The SPS method has demonstrated wide applicability: it has been used to measure the thermal conductivity and heat capacity of isotropic materials across a broad thermal conductivity range from 0.1 to 2000 W/(m·K) [16], resolve full 3D thermal conductivity tensors in anisotropic systems [20], evaluate interface thermal conductance in semiconductor stacks [17], and quantify local convective heat transfer coefficients [21].

Unlike traditional techniques such as TDTR and FDTR, SPS does not rely on phase information. This eliminates the need for phase correction and allows for robust extraction of both cross-plane thermal resistance and areal heat capacity. These advantages make SPS particularly effective for analyzing nanoscale BEOL films, where direct access is limited and structural complexity poses challenges for conventional methods.

## 3. Results and Discussion

*3.1 A Representative Case Study*

This study employs a multi-frequency modulation strategy to precisely characterize the thermal properties of multilayered samples. Figure 2(a1-c1) presents the signals measured from the #M4 sample using a fixed laser spot size of $r_0 = 14.2$ μm at modulation frequencies of 1.5 MHz, 750 kHz, and 450 kHz, respectively. The symbols denote the measured signals, while the curves represent the best-fit model prediction. For clarity, only a small portion of the signals is displayed in the logarithmic-scale plots, whereas the full square-wave heating cycle, shown in the insets, was used for best-fitting to extract the thermal properties.

The extraction of thermal parameters depends on the signals' sensitivity, which is quantified by the sensitivity coefficient, defined as

$$S_\xi = \frac{\partial \ln A_{\text{norm}}}{\partial \ln \xi} \tag{1}$$

where $A_{\text{norm}}$ is the normalized amplitude signal, and $\xi$ represents the parameter under analysis. Figure 2(a2-c2) displays the sensitivity coefficients corresponding to the signals in Fig. 2(a1-c1) for key parameters of the layers of interest: $k_{z,\text{dil}}$, $C_\text{dil}$, and $h_\text{dil}$ for the doped interface layer, as well as $k_{z,f1}$, $C_{f1}$, and $h_{f1}$ of the M1-M4 layer. The signals exhibit negligible sensitivity to $k_r$ of both layers and the interface thermal conductance $G$ between the Al transducer and the sample.

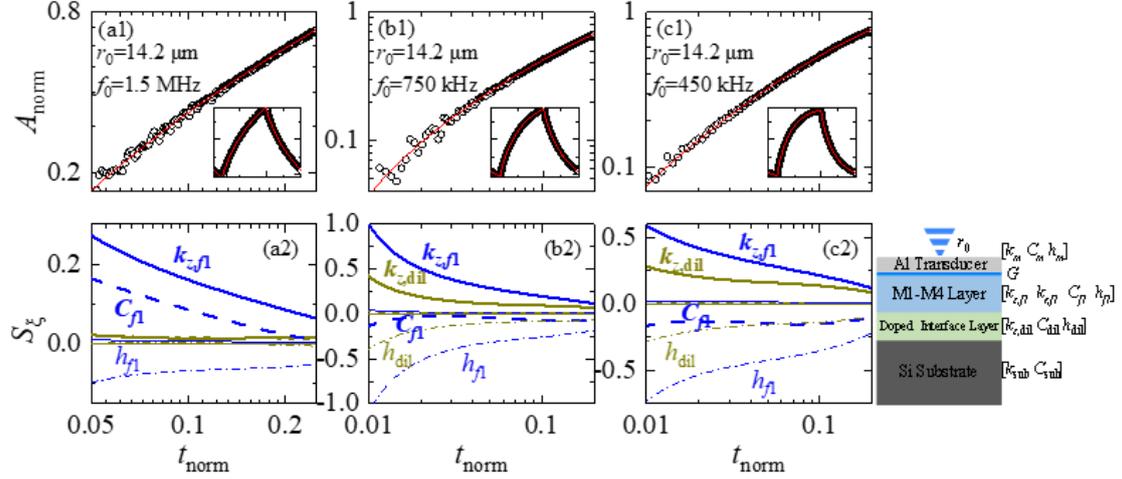

**Fig. 2** Measured signals and sensitivity analysis for the #M4 sample. (a1–c1) Thermal response signals at modulation frequencies of 1.5 MHz, 750 kHz, and 450 kHz, using a laser spot size of $r_0$=14.2 μm. Symbols represent experimental data, while curves denote best-fit model predictions. Insets display the full square-wave heating cycle. (a2–c2) Corresponding sensitivity coefficients for key thermal parameters of the doped interface layer and the M1–M4 layer.

The sensitivity analysis in Fig. 2(a2) shows that at a high modulation frequency of 1.5 MHz, which corresponds to a shallow thermal penetration depth, the measured signals are primarily sensitive to $k_{z,f1}$, $C_{f1}$, and $h_{f1}$ of the M1-M4 layer. The sensitivity coefficients satisfy $S_{h_{f1}} = S_{C_{f1}} - S_{k_{z,f1}}$, indicating that the signals are essentially sensitive to the cross-plane thermal resistance $h_{f1}/k_{z,f1}$ and the areal heat capacity $h_{f1}C_{f1}$ of the M1-M4 layer. With a known $h_{f1}$, both $k_{z,f1}$ and $C_{f1}$ can be determined by best-fitting this signal set.

As the modulation frequency decreases to 750 kHz and 450 kHz, resulting in greater thermal penetration depths, the signals remain sensitive to $h_{f1}/k_{z,f1}$ and $h_{f1}C_{f1}$ of the M1-M4 layer, while also becoming increasingly sensitive to the thermal resistance $h_\text{dil}/k_{z,\text{dil}}$ of the doped interface layer. With the properties of the M1-M4 layer pre-determined and $h_\text{dil}$ known, $k_{z,\text{dil}}$ can be extracted by best-fitting the low-frequency signal sets.

Therefore, by simultaneously fitting the three signal sets in Fig. 2, we determine $k_{z,f1} = 0.65 \pm 0.066\,\text{W/(m·K)}$ and $C_{f1} = 0.97 \pm 0.08\,\text{MJ/(m}^3\cdot\text{K)}$ for the M1-M4 layer, as well as $k_{z,\text{dil}} = 1.01 \pm 0.17 \, \text{W/(m·K)}$ for the doped interface layer. Uncertainties are estimated using a comprehensive error propagation formula that incorporates uncertainties in input parameters as well as contributions from signal noise and fitting quality. Among the input parameters, we assume uncertainties of 10% for $k_m$ and $k_{\text{sub}}$, 3% for $C_m$ and $C_{\text{sub}}$, 5% for $h_m$, $h_{f1}$, and $h_{\text{dil}}$, and 2% for $r_0$. Since the signals are insensitive to the doped interface layer's heat capacity, it is treated as an input parameter with a fixed value of 1.5 MJ/(m³·K) and an assumed uncertainty of 50%. Further details on this error propagation formula can be found in some previous literature [16].

We then perform measurements on the #M6 sample at the same location, using a similar laser spot size of $r_0 = 14.7$ μm and modulation frequencies of 1.5 MHz, 450 kHz, and 150 kHz. The corresponding measured signals are shown in Fig. 3(a1-c1). Due to slight variations in optical alignment between measurements on different samples, the laser spot size in this set of measurements differs slightly from that in Fig. 2. However, we carefully calibrate the laser spot size for each measurement to ensure high accuracy.

We first analyze the measured signals from the #M6 sample using a two-layer model comprising the doped interface layer and the M1-M6 layer. As shown in Fig. 3(a2) and (b2), at modulation frequencies above 450 kHz, the signals are sensitive only to the M1-M6 layer, with negligible sensitivity to the doped interface layer. This lower threshold frequency, compared to the #M4 sample, results from the increased thickness of the M1-M6 layer.

Above 450 kHz, the signals are primarily sensitive to $h_f/k_{z,f}$ and $h_f C_f$ of the M1-M6 layer. While sensitivity to $h_f/k_{z,f}$ remains largely unchanged, sensitivity to $h_f C_f$ increases significantly as frequency decreases. Therefore, by fitting signals at 450 kHz and 1.5 MHz, both $k_{z,f}$ and $C_f$ can be determined, given a known $h_f$.

As the modulation frequency decreases to 150 kHz, the signals also become sensitive to $h_{\text{dil}}/k_{z,\text{dil}}$ of the doped interface layer. With the properties of the M1-M6 layer pre-determined and $h_{\text{dil}}$ known, $k_{z,\text{dil}}$ can be extracted. By simultaneously fitting the three signal sets in Fig. 3, we determine $k_{z,f} = 0.64 \pm 0.09 \, \text{W/(m·K)}$ and $C_f = 1.34 \pm 0.11 \, \text{MJ/(m}^3\text{·K)}$ for the M1-M6 layer, as well as $k_{z,\text{dil}} = 1.01 \pm 0.4 \, \text{W/(m·K)}$ for the doped interface layer.

Notably, the independently determined $h_{\text{dil}}/k_{z,\text{dil}}$ for the #M6 sample matches that of the #M4 sample, albeit with greater uncertainty due to the thicker M1-M6 layer. This consistency confirms the identical nature of the two samples and reinforces the reliability of the current measurements.

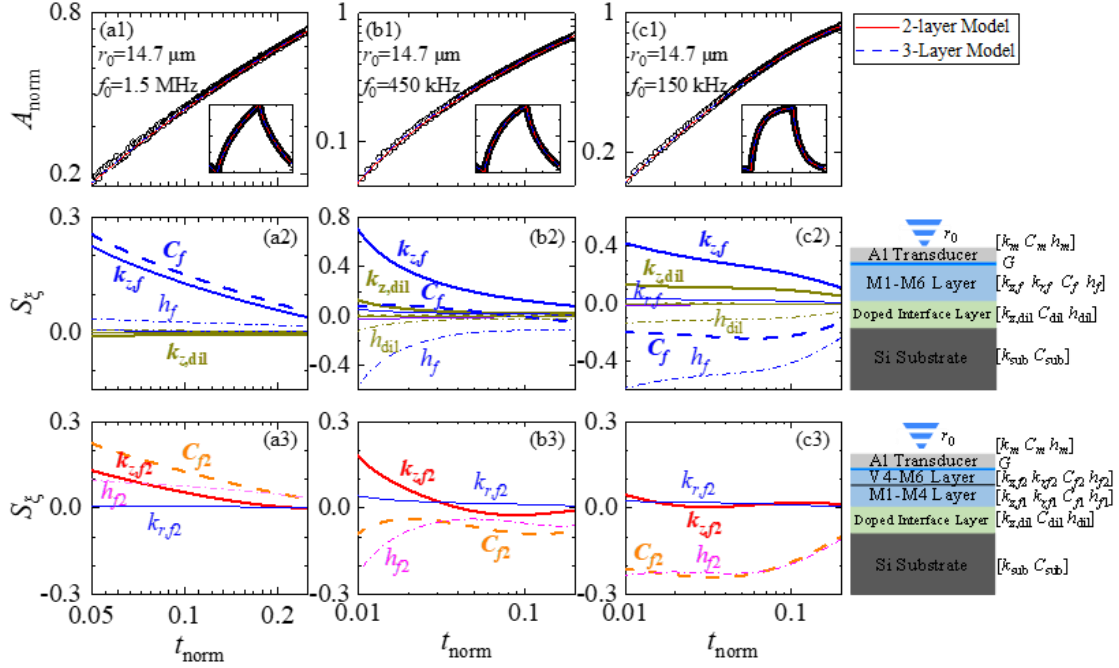

**Fig. 3** Measured signals and sensitivity analysis for the #M6 sample. (a1–c1) Thermal response signals at modulation frequencies of 1.5 MHz, 450 kHz, and 150 kHz, measured with a laser spot size of $r_0$=14.7 μm. Symbols indicate experimental data; solid and dashed curves show model predictions using two-layer and three-layer thermal models, respectively. (a2–c2) Sensitivity coefficients based on the two-layer model, highlighting key parameters of the doped interface layer and the M1-M6 layer. (a3–c3) Sensitivity coefficients based on the three-layer model, focusing on the V4-M6 sublayer.

We further analyze the same sets of signals using a three-layer model consisting of the doped interface layer, the M1-M4 layer, and the V4-M6 layer. Figure 3(a3-c3) shows the sensitivity coefficients for the parameters of the V4-M6 layer. At a relatively high frequency of 1.5 MHz, the signals are sensitive to both $h_{f2}/k_{z,f2}$ and $h_{f2}C_{f2}$ of the V4-M6 layer. However, as the frequency decreases to 150 kHz, the sensitivity to $h_{f2}/k_{z,f2}$ becomes negligible, while the sensitivity to $h_{f2}C_{f2}$ remains high. Therefore, with the properties of the doped interface layer and the M1-M4 layer pre-determined from the #M4 sample and a known $h_{f2}$, both $k_{z,f2}$ and $C_{f2}$ can be determined by fitting the three sets of signals, yielding $k_{z,f2} = 0.67 \pm 0.087$ W/(m·K) and $C_{f2} = 1.42 \pm 0.19$ MJ/(m³·K) for the V4-M6 layer. The model predictions using the three-layer model, shown as dashed curves in Fig. 3(a1-c1), are identical to those obtained using the two-layer model.

Table 1 summarizes the cross-plane thermal conductivities and heat capacities of the doped interface layer, the M1-M4 layer, the V4-M6 layer, and the M1-M6 layer measured from the #M4 and #M6 samples at the same location.

**Table 1**. Summary of Measured Cross-Plane Thermal Conductivities and Heat Capacities for Different Layers

|  | $k_z$ (W/(m·K)) | | $C$ (MJ/(m³·K)) | |
| --- | --- | --- | --- | --- |
|  | #M4 | #M6 | #M4 | #M6 |
| V4-M6 | -- | 0.67 / 0.64 | -- | 1.42 / 1.34 |
| M1-M4 | 0.65 | | 0.97 | |
| DIL | 1.01 | 1.01 | -- | -- |

Results in Table 1 suggest that the heat capacities approximately add linearly as:
$$h_1 C_1 + h_2 C_2 = (h_1 + h_2) C_{\text{cmb}} \tag{2}$$
while the cross-plane thermal resistance follows the series resistance model:
$$\frac{h_1}{k_{z1}} + \frac{h_2}{k_{z2}} = \frac{h_1 + h_2}{k_{z,\text{cmb}}} \tag{3}$$
where $k_{z,\text{cmb}}$ and $C_{\text{cmb}}$ represent the thermal conductivity and heat capacity of the composite film consisting of layers 1 and 2, respectively.

*3.2 Dependence on Metal Density*

The same measurements were repeated at 25 different locations on the samples. Figure 4 summarizes the normalized cross-plane thermal resistances $R_L/R_{L,\max}$ and areal heat capacities $((h_f C_f)/(h_f C_f)_{\max})$ of the composite film comprising layers from M1 to M6, as functions of the average metal density of the M and V layers ($\langle \rho_M \rangle / \langle \rho_M \rangle_{\max}$ and $\langle \rho_V \rangle / \langle \rho_V \rangle_{\max}$), respectively. The average metal density of the M layers is defined as

$$\langle \rho_M \rangle = \frac{1}{N} \sum_{i=1}^{N} \rho_{M_i} \tag{4}$$

where $\rho_{M_i}$ is the metal density of the *i*-th metal interconnection layer within a 50 μm × 50 μm area centered around the laser spot, obtained from the graphic database system (GDS) file, and $N$ is the total number of M layers in the analyzed composite film. The definition of the average metal density of the V layers, $\langle\rho_V\rangle$, follows a similar approach.

Values calculated from the summation of the M1–M4 and V4–M6 layers (presented as solid symbols in the plots) exhibit good agreement with the values calculated based on the effective thermal conductivity and heat capacity of the M1–M6 layer (open symbols), confirming that the total thermal resistance follows the series resistance model, while the areal heat capacity follows the simple summation rule. These data are further presented in contour maps in Fig. 4(a3) and (b3) as functions of $\langle\rho_M\rangle/\langle\rho_M\rangle_{max}$ and $\langle\rho_V\rangle/\langle\rho_V\rangle_{max}$, showing that $R_L$ decreases and $hC$ increase with both $\langle\rho_M\rangle$ and $\langle\rho_V\rangle$.

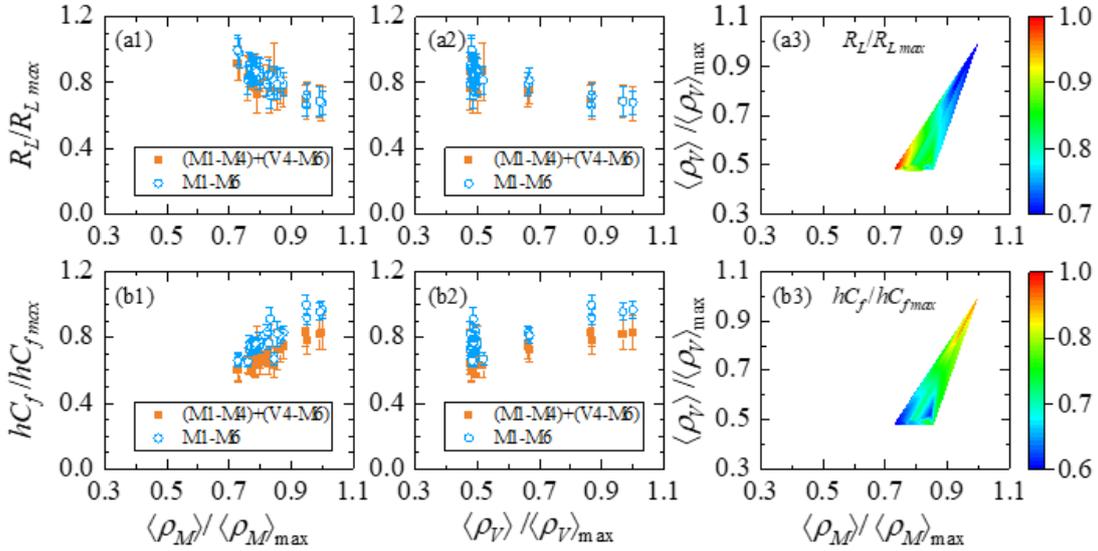

**Fig. 4** Dependence of normalized cross-plane thermal resistance and areal heat capacity on average metal densities in the BEOL stack.
(a1, a2) Normalized thermal resistances $R_L/R_{L,max}$ of the M1-M6 layer as functions of normalized average metal densities in the interconnect ($\langle\rho_M\rangle/\langle\rho_M\rangle_{max}$) and via ($\langle\rho_V\rangle/\langle\rho_V\rangle_{max}$) layers, respectively.
(b1, b2) Normalized areal heat capacities $(h_f C_f)/(h_f C_f)_{max}$ as functions of the same variables. Solid symbols represent summed contributions from the M1-M4 and V4-M6 layers; open symbols show direct measurements of the full M1-M6 layer.
(a3, b3) Contour maps showing how thermal resistance and areal heat capacity vary with metal density in the interconnect and via layers.

*3.3 COMSOL Simulation Analysis*

To mechanistically interpret the experimental results and identify the physical drivers of cross-plane heat transport in BEOL stacks, we developed a finite element model using COMSOL Multiphysics. The simulations systematically investigated the impact of three key factors on the effective cross-plane thermal conductivity ($k_{\text{eff}}$): (1) dielectric thermal conductivity ($k_d$), (2) via connectivity, and (3) copper-dielectric interfacial thermal conductance ($G$).

*3.3.1 Model configuration*

We simulated a representative BEOL unit cell with dimensions of 1 μm × 1 μm × 400 nm, replicating the M1-M4 multilayer structure of the #M4 sample (see Fig. 1a). Periodic boundary conditions were applied laterally to emulate an infinite array of repeating units. A fixed temperature of 300 K was applied at the bottom (M1 layer), while a constant heat flux of $3\times10^7$ W/m² was imposed at the top (M4 layer), reflecting experimental conditions. The effective thermal conductivity $k_{\text{eff}}$ was determined under steady-state conditions using:

$$k_{\text{eff}} = \frac{Q \cdot h}{\Delta T} \quad (5)$$

where $Q$ is the applied heat flux, $h$ is the stack thickness, and $\Delta T$ is the average temperature difference between M1 and M4 layers.

The thermal conductivity of copper interconnects was set to $k_{\text{Cu}} = 100$ W/(m·K), accounting for size effects in nanoscale Cu lines [22]. The dielectric thermal conductivity $k_d$ was varied from 0.2 to 0.8 W/(m·K), representing the range of typical low-$k$ materials (Fig. 5, data from [23-28]). The interfacial thermal conductance $G$ between Cu and dielectrics was swept from 5 to 50 MW/(m²·K), consistent with metal-dielectric interfaces incorporating etch-stop layers [29, 30].

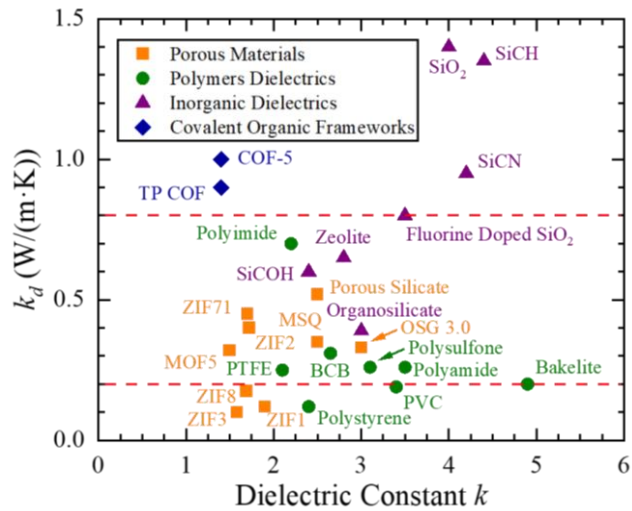

**Fig. 5** Literature survey [23-28] of thermal conductivity vs dielectric constant for low-$k$ materials.

To assess the influence of via connectivity, we modeled three via designs, as shown in Fig. 6(a1-c1):

1) High-conductivity configuration, with vias aligned vertically to form direct thermal pathways;
2) Random configuration, with vias distributed stochastically;
3) Low-conductivity configuration, with vias arranged in tortuous, disconnected paths to maximize thermal resistance.

All models used identical material properties and metal densities to isolate the effect of connectivity.

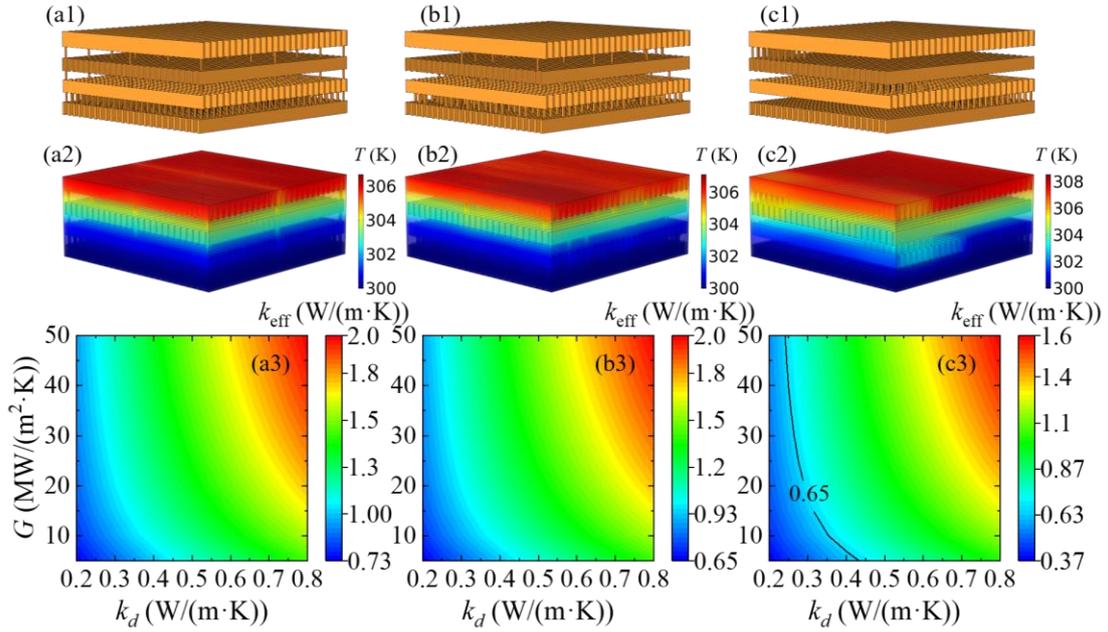

**Fig. 6** Temperature distributions and effective thermal conductivity in BEOL unit cells with varying via connectivity.
(a1–c1) Schematics of three via configurations: high-conductivity (direct paths), random, and low-conductivity (tortuous, disconnected paths).
(a2–c2) Simulated temperature fields for three via configurations under identical boundary conditions.
(a3–c3) Corresponding contour maps of effective thermal conductivity ($k_{eff}$) as functions of dielectric thermal conductivity ($k_d$) and interfacial thermal conductance ($G$) for each configuration. A black contour line in (c2) indicates the experimentally measured value of $k_{eff}$=0.65 W/(m·K), suggesting the closest match to the low-conductivity model.

*3.3.2 Results and discussion*

Temperature distributions in Fig. 6(a2-c2) reveal that disrupted via paths in the low-conductivity model create localized hot spots. In contrast, the high-conductivity model demonstrates the most effective heat spreading. As shown in Fig. 6(a3-c3), the high-conductivity configuration improves $k_{\text{eff}}$ by up to 62% (0.65 W/(m·K) vs. 1.05 W/(m·K)) relative to the low-conductivity design. The random configuration yields intermediate performance, confirming that via alignment significantly impacts thermal transport.

Among all parameters studied, $k_d$ excerts the greatest influence on $k_{\text{eff}}$ due to the dielectrics' high volume fraction (65%–83% in interconnect layers and over 97% in via layers). In the high-conductivity model, increasing $k_d$ from 0.2 to 0.8 W/(m·K) enhances $k_{\text{eff}}$ by 102% (Fig. 6(a3)), with similar trends observed in other configurations. This dominance of dielectric properties underscores their critical role in determining cross-plane heat transport.

While via connectivity matters, its impact is secondary to that of bulk dielectric conductivity. Even in optimally connected configurations, heat flow remains constrained by the surrounding low-k matrix. Similarly, the effect of interfacial thermal conductance $G$ shows saturation behavior—$k_{\text{eff}}$ increases with $G$ up to ~30 MW/(m²·K) but plateaus beyond that (Fig. 6(a3-c3)), suggesting that further interfacial optimization yields diminishing returns.

*3.3.3 Design implications*

These results establish a clear hierarchy in thermal design priorities for BEOL stacks. First and foremost is the material-centric optimization. Enhancing dielectric thermal conductivity offers the most substantial performance gains. Materials with $k_d >$ 0.5 W/(m·K), such as COFs [23], are particularly promising. The second consideration is structural enhancement. Via alignment should prioritize vertical continuity. Even random distributions outperform tortuous pathways, indicating that via connectivity should not be neglected. Interface engineering comes last. The Cu-dielectric interfacial thermal conductance should be engineered to exceed 30 MW/(m²·K) but need not be excessively optimized beyond this point.

In summary, COMSOL simulations reveal that dielectric thermal conductivity is the primary lever for improving BEOL thermal performance, followed by via connectivity and interfacial thermal conductance. The high-conductivity configuration consistently achieves superior $k_{\text{eff}}$ across all parameter ranges, emphasizing the importance of designing efficient vertical heat conduction paths. These insights provide practical guidance for thermal-aware design of advanced ICs, highlighting the need for coordinated improvements in material selection and structural architecture for effective heat management in 3D-stacked devices.

## 4. Conclusions

This study establishes the SPS method as a robust, high-resolution, and phase-free technique for quantitatively characterizing cross-plane thermal transport in BEOL multilayer structures. By applying SPS to two semiconductor chips polished to different interconnect levels (M4 and M6), we show that cross-plane thermal resistance follows a series resistance model, while areal heat capacity scales linearly with metal content. These experimentally validated trends provide direct evidence of the structural dependence of thermal transport in BEOL architectures.

To gain deeper physical insight, we performed finite element simulations using COMSOL Multiphysics, systematically evaluating the effects of dielectric thermal conductivity, via connectivity, and interfacial thermal conductance. These results establish a clear hierarchy among these factors. Dielectric thermal conductivity represents the most significant influence due to the high volume fraction of dielectrics in BEOL stacks. Increasing dielectric thermal conductivity from 0.2 to 0.8 W/(m·K) more than doubles the effective thermal conductivity. Via connectivity plays a secondary but non-negligible role, with well-aligned vias enhancing cross-plane heat transport by up to 62% compared to tortuous paths. Interfacial thermal conductance offers marginal improvements beyond 30 MW/(m²·K), indicating diminishing returns at higher interface quality.

Our integrated approach not only enables precise extraction of local thermal parameters but also reveals the mechanistic interplay between materials and geometry in BEOL heat conduction. These insights offer actionable design strategies for next-generation, thermally optimized semiconductor devices and establish the SPS method as a valuable tool for guiding the thermal design of complex multilayer systems.

**DATA AVAILABILITY**

The data that support the findings of this study are within the main text.

**DECLARATION OF COMPETING INTEREST**

The authors declare that they have no known competing financial interests or personal relationships that could have appeared to influence the work reported in this paper.

**ACKNOWLEDGMENT**

This work is supported by the National Natural Science Foundation of China (NSFC) through Grant No. 52376058.

**REFERENCES**

[1] P. Benkart, A. Kaiser, M. Bschorr, H. Huebner, A. Heittmann, A. Munding, H. Pfleiderer, E. Kohn, U. Ramacher, 3D Chip Stack Technology Using Through-Chip Interconnects, IEEE Design & Test of Computers, 22 (2005) 512-518.

[2] K. Cao, J. Zhou, T. Wei, M. Chen, S. Hu, K. Li, A survey of optimization techniques for thermal-aware 3D processors, Journal of Systems Architecture, 97 (2019) 397-415.

[3] X. Zhou, J. Yang, Y. Xu, Y. Zhang, J. Zhao, Thermal-Aware Task Scheduling for 3D Multicore Processors, IEEE Transactions on Parallel and Distributed Systems, 21 (2010) 60-71.

[4] V. d'Alessandro, A. Magnani, L. Codecasa, N. Rinaldi, K. Aufinger, Advanced thermal simulation of SiGe:C HBTs including back-end-of-line, Microelectronics Reliability, 67 (2016).

[5] S. Kikuchi, M. Suwada, H. Onuki, Y. Iwakiri, N. Nakamura, Thermal characterization and modeling of BEOL for 3D integration, 2015 IEEE CPMT Symposium Japan (ICSJ), DOI (2015) 97-100.

[6] N. Nakamura, Y. Iwakiri, H. Onuki, M. Suwada, S. Kikuchi, Thermal modeling and experimental study of 3D stack package with hot spot consideration, 2015 IEEE 65th Electronic Components and Technology Conference (ECTC), 2015, pp. 1742-1748.

[7] X. Chang, H. Oprins, M. Lofrano, B. Vermeersch, I. Ciofi, Z. Tokei, I. De Wolf, Thermal analysis of advanced back-end-of-line structures and the impact of design parameters, 2022.

[8] X. Chang, H. Oprins, B. Vermeersch, V. Cherman, M. Lofrano, S. Park, Z. Tokei, I.D. Wolf, Experimentally Validated Thermal Modeling Prediction for BEOL and BSPDN Stacks, 2024 IEEE 74th Electronic Components and Technology Conference (ECTC), 2024, pp. 498-505.

[9] J. Choi, U. Monga, Y. Park, H. Shim, U. Kwon, S. Pae, D.S. Kim, Impact of BEOL Design on Self-heating and Reliability in Highly-scaled FinFETs, 2019 International Conference on Simulation of Semiconductor Processes and Devices (SISPAD), 2019, pp. 1-4.

[10] E.G. Colgan, R.J. Polastre, J. Knickerbocker, J. Wakil, J. Gambino, K. Tallman, Measurement of back end of line thermal resistance for 3D chip stacks, 29th IEEE Semiconductor Thermal Measurement and Management Symposium, 2013, pp. 23-28.

[11] P. Jiang, X. Qian, R. Yang, Tutorial: Time-domain thermoreflectance (TDTR) for thermal property characterization of bulk and thin film materials, J. Appl. Phys., 124 (2018) 161103.

[12] A.J. Schmidt, PUMP-PROBE THERMOREFLECTANCE, Annu. Rev. Heat Transfer, 16 (2014) 159.

[13] C. Wei, X. Zheng, D.G. Cahill, J.C. Zhao, Invited article: micron resolution spatially resolved measurement of heat capacity using dual-frequency time-domain thermoreflectance, Rev. Sci. Instrum., 84 (2013) 071301.

[14] E. Ziade, Wide bandwidth frequency-domain thermoreflectance: Volumetric heat capacity,

anisotropic thermal conductivity, and thickness measurements, Rev. Sci. Instrum., 91 (2020) 124901.

[15] A.J. Schmidt, R. Cheaito, M. Chiesa, A frequency-domain thermoreflectance method for the characterization of thermal properties, Rev. Sci. Instrum., 80 (2009) 094901.

[16] T. Chen, S. Song, Y. Shen, K. Zhang, P. Jiang, Simultaneous measurement of thermal conductivity and heat capacity across diverse materials using the square-pulsed source (SPS) technique, International Communications in Heat and Mass Transfer, 158 (2024) 107849.

[17] S. Song, T. Chen, P. Jiang, Comprehensive thermal property measurement of semiconductor heterostructures using the square-pulsed source (SPS) method, J. Appl. Phys., 137 (2025) 055101.

[18] T. Chen, P.-Q. Jiang, Unraveling intrinsic relationship of thermal properties in thermoreflectance experiments, Acta Physica Sinica, 73 (2024) 230202.

[19] T. Chen, P. Jiang, Decoupling Thermal Properties in Multilayer Systems for Advanced Thermoreflectance Techniques, arXiv preprint arXiv:2410.08480, DOI (2024).

[20] T. Chen, S. Song, R. Hu, P. Jiang, Comprehensive measurement of three-dimensional thermal conductivity tensor using a beam-offset square-pulsed source (BO-SPS) approach, International Journal of Thermal Sciences, 207 (2025) 109347.

[21] T. Chen, P. Jiang, An optical method for measuring local convective heat transfer coefficients over 100 W/(m2·K) with sub-millimeter resolution, Measurement, 249 (2025) 117001.

[22] X. Chang, B. Vermeersch, H. Oprins, M. Lofrano, V. Cherman, S. Park, Z. Tokei, I.D. Wolf, Thermal Modeling and Analysis of Equivalent Thermal Properties for Advanced BEOL Stacks, IEEE Transactions on Components, Packaging and Manufacturing Technology, DOI 10.1109/TCPMT.2025.3564833(2025) 1-1.

[23] A.M. Evans, A. Giri, V.K. Sangwan, S. Xun, M. Bartnof, C.G. Torres-Castanedo, H.B. Balch, M.S. Rahn, N.P. Bradshaw, E. Vitaku, D.W. Burke, H. Li, M.J. Bedzyk, F. Wang, J.-L. Brédas, J.A. Malen, A.J.H. McGaughey, M.C. Hersam, W.R. Dichtel, P.E. Hopkins, Thermally conductive ultra-low-k dielectric layers based on two-dimensional covalent organic frameworks, Nature Materials, 20 (2021) 1142-1148.

[24] M.T. Alam, R.A. Pulavarthy, J. Bielefeld, S.W. King, M.A. Haque, Thermal Conductivity Measurement of Low-k Dielectric Films: Effect of Porosity and Density, J. Electron. Mater., 43 (2014) 746-754.

[25] B. Santhosh, M. Biesuz, G.D. Sorarù, Thermal properties of dense polymer-derived SiCN(O) glasses, Mater. Lett., 288 (2021) 129336.

[26] X. Xu, Z. Wang, Thermal Conductivity Enhancement of Benzocyclobutene With Carbon Nanotubes for Adhesive Bonding in 3-D Integration, IEEE Transactions on Components, Packaging and Manufacturing Technology, 2 (2012) 286-293.

[27] H. Oprins, V. Cherman, B. Vermeersch, F. Luciano, X. Chang, V. Founta, Y. Ding, C. Adelmann, Z. Tokei, Experimental thermal characterization of thin film low-k dielectric materials,  2024 23rd IEEE Intersociety Conference on Thermal and Thermomechanical Phenomena in Electronic Systems (ITherm), IEEE, 2024, pp. 1-8.

[28] D. Xi-Jie, H. Yi-Fan, W. Yu-Ying, Z. Jun, W. Zhen-Zhu, A fractal model for effective thermal conductivity of isotropic porous silica low-k materials, Chinese Physics Letters, 27 (2010) 044401.

[29] J.C. Duda, T.E. Beechem, J.L. Smoyer, P.M. Norris, P.E. Hopkins, Role of dispersion on phononic thermal boundary conductance, J. Appl. Phys., 108 (2010) 073515.

[30] R.J. Stevens, A.N. Smith, P.M. Norris, Measurement of Thermal Boundary Conductance of a Series of Metal-Dielectric Interfaces by the Transient Thermoreflectance Technique, J. Heat Transfer, 127 (2005)

315.